\documentclass[lettersize,hidelinks,journal]{IEEEtran}

\usepackage{amsmath,amssymb,amsfonts}
\usepackage{array}
\usepackage{textcomp}
\usepackage{stfloats}
\usepackage{url}
\usepackage{verbatim}
\usepackage{graphicx}
\usepackage{cite}
\usepackage{xcolor}
\usepackage{orcidlink}
\usepackage[acronym,shortcuts]{glossaries}
\usepackage{bm}
\usepackage[font=footnotesize]{caption}
\usepackage{subcaption}
\usepackage{relsize}
\usepackage{soul}
\usepackage{algorithm, algpseudocode}
\usepackage{algcompatible}
\usepackage{lipsum}
\usepackage{mathtools}
\usepackage[bottom,hang,flushmargin]{footmisc}
\usepackage{tabularx}

\def\BibTeX{{\rm B\kern-.05em{\sc i\kern-.025em b}\kern-.08em
T\kern-.1667em\lower.7ex\hbox{E}\kern-.125emX}}

\newcommand{\herm}[0]{^{\mathsf{H}}}

\newacronym{AFDM}{AFDM}{affine frequency division multiplexing}
\newacronym{IM}{IM}{index modulation}
\newacronym{RF}{RF}{radio frequency}
\newacronym{RIS}{RIS}{reflective intelligent surface}
\newacronym{ISAC}{ISAC}{integrated sensing and communications}
\newacronym{6G}{6G}{sixth-generation}
\newacronym{B6G}{B6G}{beyond-sixth-generation}
\newacronym{B5G}{B5G}{beyond-fifth-generation}
\newacronym{DD}{DD}{doubly-dispersive}
\newacronym{CPIM}{CPIM}{chirp-permutation-index modulation}
\newacronym{PMF}{PMF}{probability mass function}
\newacronym{GAS}{GAS}{Grover adaptive search}
\newacronym{MIMO}{MIMO}{multiple-input multiple-output}
\newacronym{MMSE}{MMSE}{minimum mean-squared-error}
\newacronym{ML}{ML}{maximum likelihood}
\newacronym{SISO}{SISO}{single-input single-output}
\newacronym{AWGN}{AWGN}{additive white Gaussian noise}
\newacronym{CP}{CP}{chirp-permuted}
\newacronym{CPP}{CPP}{chirp-periodic prefix}
\newacronym{OFDM}{OFDM}{orthogonal frequency division multiplexing}
\newacronym{AFT}{AFT}{affine Fourier transform}
\newacronym{DFT}{DFT}{discrete Fourier transform}
\newacronym{IDFT}{IDFT}{inverse discrete Fourier transform}
\newacronym{DAFT}{DAFT}{discrete affine Fourier transform}
\newacronym{IDAFT}{IDAFT}{inverse discrete affine Fourier transform}
\newacronym{BER}{BER}{bit-error-rate}
\newacronym{CDF}{CDF}{cumulative distribution function}
\newacronym{QUBO}{QUBO}{quadratic unconstrained binary optimization}
\newacronym{LTV}{LTV}{linear time-variant}
\newacronym{TVIRF}{TVIRF}{time-varying impulse response function}
\newacronym{OTFS}{OTFS}{orthogonal time frequency space}
\newacronym{MSE}{MSE}{mean-squared-error}
\newacronym{CSI}{CSI}{channel state information}
\newacronym{BPSK}{BPSK}{binary phase shift keying}
\newacronym{QPSK}{QPSK}{quadrature phase shift keying}
\newacronym{5GNR}{5GNR}{5G New Radio}
\newacronym{EbN0}{$E_b/N_0$}{energy-per-bit-to-noise-spectral-density ratio}
\newacronym{SNR}{SNR}{signal-to-noise ratio}
\newacronym{CPD}{CPD}{chirp-permutation-domain}
\newacronym{NOMA}{NOMA}{non-orthogonal multiple access}
\newacronym{OMA}{OMA}{orthogonal multiple access}
\newacronym{AI}{AI}{artificial intelligence}
\newacronym{AN}{AN}{artificial noise}
\newacronym{XR}{XR}{extended reality}
\newacronym{IoT}{IoT}{Internet-of-Things}
\newacronym{PhySec}{PhySec}{physical-layer security} 
\newacronym{SotA}{SotA}{state-of-the-art} 

\begin{document}

\title{\vspace{-0.5ex}Chirp-Permuted AFDM for Quantum-Resilient \\[-0.1ex] Physical-Layer Secure Communications \vspace{-0.5ex}}


\author{Hyeon Seok Rou\textsuperscript{\orcidlink{0000-0003-3483-7629}}\!,~\IEEEmembership{Member,~IEEE}, Giuseppe Thadeu Freitas de Abreu\textsuperscript{\orcidlink{0000-0002-5018-8174}}\!,~\IEEEmembership{Senior Member,~IEEE}.
\thanks{Hyeon Seok Rou and Giuseppe Thadeu Freitas de Abreu are with the School of Computer Science and Engineering, Constructor University, Campus Ring 1, 28759 Bremen, Germany ([hrou, gabreu]@constructor.university).}

\vspace{-5ex}}

\markboth{}%
{H. S. Rou \MakeLowercase{\textit{et al.}}: Chirp-Permuted AFDM for Quantum-Resilient Physical-Layer Secure Communications}


\maketitle

\begin{abstract}
%
%

~\!\!{\color{black}We present} a novel physical-layer secure communications scheme based on a {\color{black}chirp-permuted variant of the \ac{AFDM} waveform recently proposed for \ac{6G} systems, which ensures that eavesdroppers unaware of the correct chirp-permutation face significant challenges in signal detection, even with perfect \ac{CSI} and co-location with the legitimate user.}
The security of the proposed scheme is studied in terms of the complexity required to ﬁnd the correct permutation via classical and quantum search algorithms, {\color{black} analytically shown to be infeasible in both cases due to the factorially-scaling search space, and the probability of breach under the random-guess approach, also shown to be negligible.}
%
%
\vspace{-0.1ex}
\end{abstract}


\vspace{-1.25ex}

\glsresetall

\IEEEpeerreviewmaketitle

\section{Introduction}
\label{sec:introduction}

The rapid evolution of wireless communications technologies, transitioning towards \ac{B5G} and \ac{6G} networks, is significantly increasing performances in terms of data rates, latency, and device connectivity, {\color{black}and perhaps more crucially,} enabling new functionalities such as edge computing, integrated \ac{AI}, and \ac{ISAC} \cite{Wang_CST23,Rou_TWC24}.

The penetration of these innovations into use cases such as critical mission, autonomous vehicles, and \ac{IoT} are, however, also expected to increase the exposure to {\color{black}new threats, making inherent security a must-have feature in \ac{6G} systems \cite{Nguyen_CST21}.
Besides the larger exposure due to the \ac{ISAC} paradigm, traditional secure communications systems are expected to face significant challenges with the rise of quantum computers, which can efficiently break widely-used public key cryptosystems such as the RSA and ECC \cite{Subramani_CS25} by solving previously prohibitive mathematical problems such as factoring large numbers or searching a large codebook \cite{Yukiyoshi_TQE24}, much faster than classical computers.

This has motivated the development of \textit{post-quantum} methods, such as post-quantum cryptography \cite{Bernstein_Nature17} and quantum key encapsulation \cite{MA_Access2025} and distribution \cite{Cao_CST22}, which exploit either the quantum entanglement phenomena or complex hash- and multivariate quadratic-based algorithms that remain unbreakable even under quantum-accelerated factorization and searches.
However, in addition to the associated large overheads, such methods depend on costly and still incipient technologies such as quantum computers and quantum channels, which pose a significant implementation challenge to their wider adoption and application \cite{Subramani_CS25}.}

{\color{black}In turn, \ac{PhySec} techniques have also gained much interest as an alternative to offer inherent security without the aforementioned challenges \cite{Mucchi_6GSecurity2021,Mitev_OJVT23}.
To this end, rather than relying on sophisticated encryption mechanisms or quantum hardware, \ac{PhySec} achieves quantum resilience by exploiting unique characteristics of the wireless medium -- such as fading, interference, and waveform modulation patterns -- to enable natively secure communications.

Good examples of \ac{PhySec} are \ac{AN} \cite{Zhang_CL18} and jamming approaches \cite{Pirayesh_CST22}, which seek to deteriorate the performance of eavesdroppers by adding optimized noise or interfering signals.
More recently, a new and interesting form of \ac{PhySec} has emerged, exploiting the low-cost electromagnetic structures such as \acp{RIS} to manipulate and optimize the propagation environment in favor of legitimate users, and to hinder eavesdroppers \cite{Cai_TCOM23}.
Despite their many interesting features, such methods still require additional energy and/or the deployment of dedicated hardware for their implementation, besides requiring (typically) accurate state information of the eavesdropper (such as channel gains or locations), which also challenge their feasibility.}

In light of the above, this article proposes {\color{black}an efficient \ac{PhySec}} communications scheme that achieve {\color{black}quantum-resilience, without quantum-reliance or the requirement of additional energy, hardware overhead, or any information about eavesdroppers.}
The proposed method leverages a recently discovered chirp-permuted variation \cite{Rou_Asilomar24} of the \ac{AFDM} \cite{Bemani_TWC23}, {\color{black} which retain all of the beneficial properties of the next-generation \ac{AFDM} waveform including excellent communications performance over doubly-dispersive channels,} in order to enable {\color{black}feasible post-quantum} security and resilience in \ac{B5G} and \ac{6G} systems.
The contributions of the article can be summarized as follows: 
{\color{black}
\begin{itemize}
\item A novel secure communications scheme based on \ac{CP}-\ac{AFDM} is proposed, which does not require any non-common energy, hardware, and complexity constraints.
The proposed scheme results in a completely undecodable received signal to the eavesdropper with the incorrect chirp-permutation order, even under co-located eavesdropping with perfect channel information.
\item The security of the proposed scheme is evaluated via numerical analysis and simulation, against eavesdroppers employing either classical or quantum-accelerated searches, or a random guess approach. 
It is demonstrated that the computational complexity required to find the correct permutation order (for successful detection) is practically infeasible even under quantum computing-aided eavesdropping, in addition to showing that the likelihood of data detection by chance is also infeasible.
\end{itemize}
}

\vspace{-1.5ex}
\section{System Model}
\label{sec:system}
\vspace{-0.5ex}

Consider a communications scenario with transmitter Alice, legitimate receiver Bob, and eavesdropper Eve, all equipped with single antennas.

Alice wishes to securely transmit data to Bob over the wireless channel, while ensuring that Eve is unable to decode the transmitted data.
{\color{black}The wireless channels} are modeled as doubly-dispersive channels with unique delay-Doppler profiles \cite{Rou_SPM24} {\color{black}as explained in the sequel, and it is assumed that Alice does not possess the \ac{CSI} of either Eve's or Bob's channels -- such that \ac{CSI}-based encryption, \ac{AN}, or jamming techniques are not feasible -- while} both Bob and Eve have perfect \ac{CSI} of their respective channels {\color{black}to perform data estimation}.
Later, we also consider the extreme case where Bob and Eve are co-located, such that Eve has the same \ac{CSI} as Bob, {\color{black}hence} receives the exact same signal{\color{black}\footnotemark}.

\vspace{-1ex}
\subsection{Received Signal Model over Doubly-Dispersive Channels}
\label{subsec:DD_channel}

Consider a doubly-dispersive wireless channel {\color{black}in a \ac{SISO} system} described by $P$ resolvable propagation paths, where each $p$-th scattering path induces a delay $\tau_p \in [0, \tau^\mathrm{max}]$ and Doppler shift $\nu_p \in [-\nu^\mathrm{max}, +\nu^\mathrm{max}]$ to the received signal.
Then, the received signal sampled at a frequency of $f_\mathrm{s} \triangleq \frac{1}{T_\mathrm{s}}$, can be efficiently represented {\color{black}via the} discrete circular convolutional {\color{black}channel model}{\color{black}\footnotemark} given by
\begin{equation}
\color{black}
\mathbf{r} \triangleq \mathbf{H} \mathbf{s}  + \mathbf{w} = \big( \textstyle\sum_{p=1}^{P}h_p \!\cdot\! \mathbf{\Phi}_{p} \!\cdot\! \mathbf{Z}^{f_p} \!\cdot\! \mathbf{\Pi}^{\ell_p}\big)\mathbf{s} + \mathbf{w} \in \mathbb{C}^{N \times 1},
\label{eq:matrix_received_signal}
\end{equation}
where $\mathbf{r} \!\in\! \mathbb{C}^{N \!\times\! 1}$, $\mathbf{s} \!\in\! \mathbb{C}^{N \!\times\! 1}$, and $\mathbf{w} \!\in\! \mathbb{C}^{N \!\times\! 1}$ are the discrete vectors of the received signal, transmit signal, and \ac{AWGN} signal, {\color{black}respectively},
$\ell_p \triangleq \lfloor \frac{\tau_p}{T_\mathrm{s}} \rceil \in \mathbb{N}_0$ and $f_p \triangleq \frac{N\nu_p}{f_\mathrm{s}} \in \mathbb{R}$ are the normalized integer path delay and normalized digital Doppler shift of the $p$-th propagation path.

Each $p$-th path {\color{black}within the} full circular convolutional matrix $\mathbf{H} \triangleq \sum_{p=1}^{P}h_p \mathbf{\Phi}_{p} \mathbf{Z}^{f_p} \mathbf{\Pi}^{\ell_p} \in \mathbb{C}^{N \times N}$, is parametrized by the complex channel fading coefficient $h_p \in \mathbb{C}$, the diagonal prefix phase matrix $\boldsymbol{\Phi}_p  \!\in\! \mathbb{C}^{N \times N}$ with the waveform-specific prefix phase function $\phi(n)$, given by
\begin{equation}
\color{black}
\!\!\!\mathbf{\Phi}_{p} \!\triangleq\! \mathrm{diag}\big[e^{-j2\pi \cdot \phi(\ell_p)}, \ldots, e^{-j2\pi \cdot\phi(1)}, 1, \ldots, 1\big] \!\in\! \mathbb{C}^{N \times N}\!,\!\!\!
\label{eq:CCP_phase_matrix}
\end{equation}
%
the diagonal roots-of-unity matrix $\mathbf{Z} \!\in\! \mathbb{C}^{N \times N}$ given by 
\begin{equation}
\mathbf{Z} \triangleq \mathrm{diag}\big[e^{-j2\pi\frac{(0)}{N}}, \,\ldots\,, e^{-j2\pi\frac{(N\!-\!1)}{N}}\big] \in \mathbb{C}^{N \times N},
\label{eq:Z_matrix}
\end{equation}
%
and the right-multiplying circular left-shift matrix $\mathbf{\Pi} \in \mathbb{C}^{N \!\times\! N}$.


\vspace{-1ex}
\subsection{Affine Frequency Division Multiplexing (\acs{AFDM}) Waveform}
\label{subsec:AFDM}

The recently proposed \ac{AFDM} waveform is a novel modulation scheme that is designed to provide high spectral efficiency and full diversity over doubly-dispersive channels, {\color{black}using chirp-domain subcarriers} and a {\color{black}lower modulation} complexity than its competitors \cite{Bemani_TWC23}.
\ac{AFDM} modulates a sequence of $N$ symbols, $\mathbf{x} \in \mathcal{X}^{N \times 1}$, whose elements are drawn from an $M$-ary complex digital constellation $\mathcal{X} \subset \mathbb{C}$, unto the time domain signal using the \ac{IDAFT}, \vspace{-0.25ex}
\begin{equation}
\mathbf{s}^{\mathrm{AFDM}} \triangleq \mathbf{A}^{\!-1} \mathbf{x} \in \mathbb{C}^{N \times 1},
\label{eq:AFDM_modulation} \vspace{-0.5ex}
\end{equation}
where the $N$-point forward \ac{DAFT} matrix $\mathbf{A}\in\mathbb{C}^{N \times N}$ and its inverse $\mathbf{A}^{-1}\in\mathbb{C}^{N \times N}$ are efficiently described {\color{black}in terms of} a \ac{DFT} matrix and two diagonal chirp sequences, as
\begin{subequations}
\begin{gather}
\mathbf{A} \triangleq \mathbf{\Lambda}_{c_2} \mathbf{F}_N \mathbf{\Lambda}_{c_1} \in \mathbb{C}^{N \times N}, 
\label{eq:DAFT_matrix}\\[0.5ex]
\mathbf{A}^{\!-1} \!\triangleq\! (\mathbf{\Lambda}_{c_2} \mathbf{F}_N \mathbf{\Lambda}_{c_1})^{-1} \!=\!  \mathbf{\Lambda}_{c_1}\herm \mathbf{F}_N\herm  \mathbf{\Lambda}_{c_2}\herm \!=\! \mathbf{A}\herm \in \mathbb{C}^{N \times N},
\label{eq:IDAFT_matrix}
\end{gather}
\label{eq:DAFT_matrices}%
\end{subequations}
\noindent where $\mathbf{F}_N \times \mathbb{C}^{N \times N}$ is the normalized $N$-point \ac{DFT} matrix, and $\mathbf{\Lambda}_{c_1} \triangleq \mathrm{diag}(\boldsymbol{\lambda}_{c_1}) \in \mathbb{C}^{N \times N}$, $\mathbf{\Lambda}_{c_2} \triangleq \mathrm{diag}(\boldsymbol{\lambda}_{c_2}) \in \mathbb{C}^{N \times N}$ are diagonal chirp matrices whose elements are described by the chirp vector $\boldsymbol{\lambda}_{c_i} \triangleq [e^{-j2\pi c_i (0)^2}, \cdots, e^{-j2\pi c_i (N-1)^2}] \in \mathbb{C}^{N \times 1}$ with a central digital frequency of $c_i$.

\footnotetext[1]{{\color{black}This models the case of a physical wiretapper with access to Bob's hardware.}}
\footnotetext[2]{For the sake of conciseness, we refer readers to the full derivation and details of the doubly-dispersive channel model to \cite{Rou_SPM24}.}


{\color{black}We clarify here that the first chirp sequence $\mathbf{\Lambda}_{c_1}$ must be optimized to ensure the full diversity characteristics of the \ac{AFDM} waveform over doubly-dispersive channels \cite{Bemani_TWC23}, while the second chirp sequence $\mathbf{\Lambda}_{c_2}$ can be manipulated without deteriorating the communications performance, and therefore is exploited in the proposed \ac{PhySec} scheme in the following.}

\vspace{-0.15ex}
\section{Proposed Secure Communications Scheme \\ via Chirp-Permuted AFDM}
\label{sec:proposed}

Given the above, we propose to permute the second chirp sequence $\mathbf{\Lambda}_{c_2}$ in the \ac{AFDM} modulation and demodulation processes (\textit{i.e.,} the \ac{DAFT} and \ac{IDAFT}), which will be shown to retain the full diversity delay-Doppler characteristics of the \ac{AFDM}, but yield two completely different physical waveforms.
This approach was leveraged in \cite{Rou_Asilomar24} to enable \ac{IM} over the \ac{AFDM} using the chirp-permutation domain codebook, and in this article, will be leveraged to provide physical-layer security against eavesdroppers.

\vspace{-1.5ex}
\subsection{Chirp-Permuted AFDM Waveform}
\label{subsec:CP-AFDM}

Let the permutation {\color{black}operation} of an arbitrary sequence of length $N$ with order $i \in \{1,\ldots, N!\}$ be denoted by $\mathrm{perm}(\,\cdot\,, \, i)$, and the permutation order $i$ is assumed to be in ascending order of the sequence element indices\footnote{\color{black}To clarify, $\mathrm{perm}([x_1,x_2,x_3], \, 1) = [x_1,x_2,x_3]$, $\mathrm{perm}([x_1,x_2,x_3], \, 2)$ $= [x_1,x_3,x_2]$, $\mathrm{perm}([x_1,x_2,x_3], \,3)$ $= [x_2, x_1, x_3]$, \textit{etc.}}.
Then, by defining the diagonal matrix formed from the $i$-th permutation of the second chirp sequence $\boldsymbol{\lambda}_{c_2}$ as
\begin{equation}
\mathbf{\Lambda}_{c_2,i} \triangleq \mathrm{diag}\big(\mathrm{perm}(\boldsymbol{\lambda}_{c_2},i)\big) \in \mathbb{C}^{N \times N},
\label{eq:chirp_permutation}
\end{equation}
the $i$-th permuted \ac{DAFT} and \ac{IDAFT} matrices are given by
\begin{gather}
\color{black}
\!\!\!\mathbf{A}_{i} \triangleq \mathbf{\Lambda}_{c_2,i} \mathbf{F}_N \mathbf{\Lambda}_{c_1} ~\text{and}~
\mathbf{A}_{i}^{-1} \triangleq  \mathbf{\Lambda}_{c_1}\herm \mathbf{F}_N\herm \mathbf{\Lambda}_{c_2,i}\herm \in \mathbb{C}^{N \!\times\! N}, \!\!\\[-4ex] \nonumber
\end{gather} 
which is still a linear unitary transform.

In all, the chirp-permuted \ac{AFDM} waveform with the $i$-th permutation order{\color{black}\footnote{\color{black}A chirp-permuted \ac{AFDM} system with a given fixed permutation order trivially achieves the same spectral efficiency as a conventional \ac{AFDM} \cite{Bemani_TWC23}.}} is given by
\begin{equation}
\mathbf{s}_i \triangleq \mathbf{A}_{i}^{-1}\mathbf{x} = \mathbf{\Lambda}_{c_1}\herm \mathbf{F}_N\herm \mathbf{\Lambda}_{c_2,i}\herm \mathbf{x} \in \mathbb{C}^{N \times 1}.
\vspace{-0.5ex}
\end{equation}

\vspace{-2.2ex}
\subsection{Effective Channel Analysis of the Chirp-Permuted AFDM}
\label{subsec:eff_channel}

{\color{black}The} effective channel of the chirp-permuted \ac{AFDM} waveform representing the input-output relationship of the transmit symbols $\mathbf{x}$, can be obtained {\color{black}under} the demodulation with the matched permuted-\ac{DAFT} matrix $\mathbf{A}_i$, which is given by
\begin{align}
\mathbf{G}_i &\triangleq \mathbf{A}_i\mathbf{H}\mathbf{A}^{-1}_i = \big(\mathbf{\Lambda}_{c_2,i} \mathbf{F}_N \mathbf{\Lambda}_{c_1}\big) \mathbf{H} \big(\mathbf{\Lambda}_{c_1}\herm \mathbf{F}_N\herm \mathbf{\Lambda}_{c_2,i}\herm\big) \nonumber \\
&= \mathbf{\Lambda}_{c_2,i}\mathbf{\Xi}\mathbf{\Lambda}_{c_2,i}\herm ~\in \mathbb{C}^{N \times N}, \label{eq:permutation_matchedAFDM}\\[-3.5ex]\nonumber
\end{align}
where ${\mathbf{\Xi}} \triangleq \mathbf{F}_N \mathbf{\Lambda}_{c_1} \mathbf{H}\mathbf{\Lambda}_{c_1}\herm \mathbf{F}_N\herm \in \mathbb{C}^{N \times N}$ denotes the intermediate effective channel before the second chirp domain.

\begin{figure}[H]
\centering
\includegraphics[width=0.95\columnwidth]{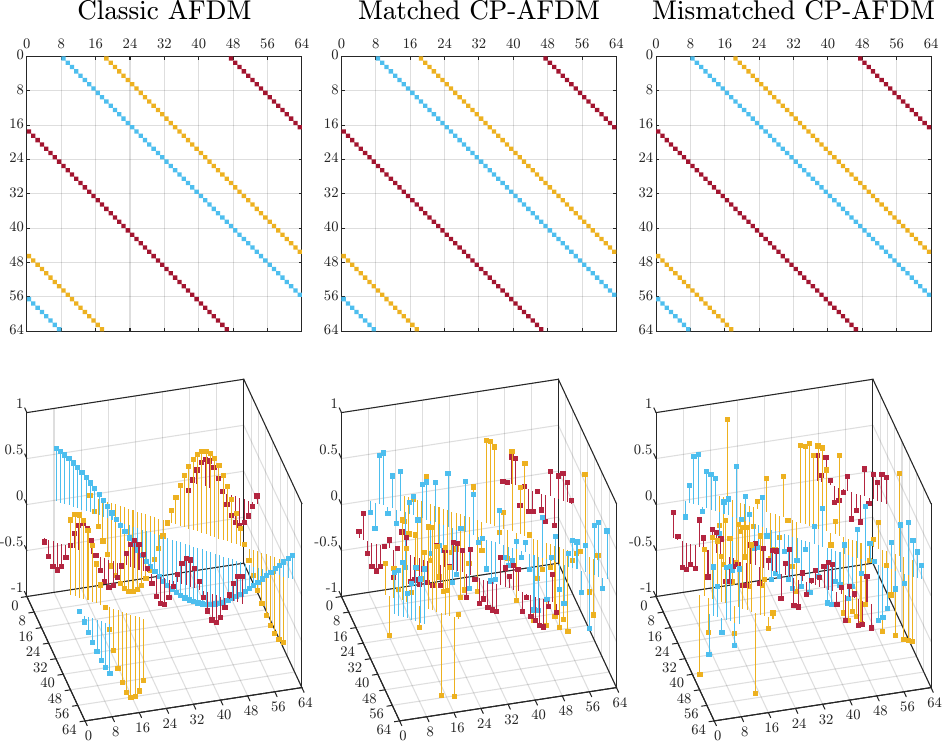}
\vspace{-0.5ex}
\caption{Illustrations of the effective channels {\color{black}$\mathbf{G} \!\in\! \mathbb{C}^{N \!\times\! N}\!$ with $N \!=\! 64$,} of the \ac{AFDM} \cite{Rou_SPM24}, \ac{CP}-\ac{AFDM} \eqref{eq:permutation_matchedAFDM}, and mismatched \ac{CP}-\ac{AFDM} \eqref{eq:y_eve}, respectively in 2D (highlighting \textbf{structure}) and 3D (highlighting \textbf{coefficients}). {\color{black} The 2D axes denote the matrix indices of the effective channels $\mathbf{G}$, and the vertical axis in 3D denotes the normalized magnitude of the non-zero channel coefficients.}}
\label{fig:eff_channel}
\vspace{-1ex}
\end{figure}

The above formulation reveals an important property of chirp-permuted \ac{AFDM} signals, and for \ac{AFDM} waveforms in general\footnote{The \ac{CP}-\ac{AFDM} reduces to the classical \ac{AFDM} of \cite{Bemani_TWC23} with $i = 1$.}.
As the second chirp (pre-chirp domain) operations are performed by diagonal matrices $\mathbf{\Lambda}_{c_2}$ and $\mathbf{\Lambda}_{c_2}\herm$, the induced effects are {\color{black} only element-wise scaling and hence a similarity transformation of the intermediate effective channel $\mathbf{\Xi}$, as seen in \eqref{eq:permutation_matchedAFDM}.}
In other words, {\color{black}the pre-chirp operations \ul{do not change the structure} of the effective channel (i.e., the positions of the non-zero elements in $\mathbf{G}$)} and therefore does not effect on the delay-Doppler orthogonality properties of the \ac{AFDM}.

This behavior is clearly illustrated in Figure \ref{fig:eff_channel}, where the chirp-permuted \ac{AFDM} waveforms result in identical channel structures as the classic \ac{AFDM}{\color{black}, as seen by the same non-zero positions of the matrix viewed in 2D}, and hence retains the same delay-Doppler characteristics -- but results in vastly different channel coefficients{\color{black}, as seen by the completely different channel coefficients viewed in 3D.
It should be highlighted that the channel coefficients are also completely different between the matched and non-matched chirp-permuted \ac{AFDM} systems, and hence yield different received signals which is the basis of the proposed scheme as detailed in the following.}

\vspace{-1.1ex}
\subsection{Secure Communications Scheme}
\label{subsec:proposed_scheme}
\vspace{-0.5ex}

Finally, we propose the secure communications scheme based on the above-described \ac{CP}-\ac{AFDM} waveform, which enables quantum-resilient physical-layer security against eavesdroppers in a doubly-dispersive environment.

Assume that Alice and Bob share a predetermined secret key $k \in \{1, \ldots, N!\}$, unknown to Eve, which is used to determine the permutation order of the chirp sequence in the \ac{CP}-\ac{AFDM} waveform.
Alice transmits the data symbol vector $\mathbf{x} \in \mathbb{C}^{N \times 1}$ modulated via the \ac{CP}-\ac{AFDM} signal $\mathbf{s}_k \triangleq \mathbf{A}_k^{\!-1} \mathbf{x} \in \mathcal{X}^{N \times 1}$ with the secret chirp-permutation order $k$.
The corresponding received signal at Bob and Eve are given by \vspace{-0.5ex}
\begin{subequations}
\begin{align}
\mathbf{r}_{\mathrm{B}} \triangleq \mathbf{H}_\mathrm{B} \mathbf{A}_k^{\!-1} \mathbf{x} + \mathbf{w}_\mathrm{B} \in \mathbb{C}^{N \times 1}, \label{eq:r_bob}\\
\mathbf{r}_{\mathrm{E}} \triangleq \mathbf{H}_\mathrm{E} \mathbf{A}_k^{\!-1} \mathbf{x} + \mathbf{w}_\mathrm{E} \in \mathbb{C}^{N \times 1}, \label{eq:r_eve} \\[-3.75ex] \nonumber
\end{align}
\label{eq:r_both}
\end{subequations}

\vspace{-3ex}
\noindent where $\mathbf{w}_\mathrm{B} \sim \mathcal{CN}(\mathbf{0},\sigma^2_\mathrm{B}\mathbf{I}_N)$ and $\mathbf{w}_\mathrm{E} \sim \mathcal{CN}(\mathbf{0},\sigma^2_\mathrm{E}\mathbf{I}_N)$ 
are the \ac{AWGN} vectors respectively at Bob and Eve, with corresponding noise variances $\sigma^2_{\mathrm{B}}$ and $\sigma^2_{\mathrm{E}}$.

It is assumed that both Bob and Eve have perfect \ac{CSI} of their respective channels, $\mathbf{H}_\mathrm{B} \in \mathbb{C}^{N \times N}$ and $\mathbf{H}_\mathrm{E} \in \mathbb{C}^{N \times N}$, such that they can perform perfect equalization to decode the transmitted data.
However, recall that only Bob has the correct permutation key $k \in \{1,\ldots,N!\}$ and therefore the correct demodulator $\mathbf{A}_k$, while Eve does not have the correct permutation order, and must use some arbitrary key $k' \in \{1,\ldots,N!\}$ and the corresponding demodulator $\mathbf{A}_{k'}$.

The demodulated signals are correspondingly given by
\begin{subequations}
\begin{align}
\mathbf{y}_\mathrm{B} &\triangleq \mathbf{A}_k \mathbf{r}_\mathrm{B} \;= \mathbf{A}_k \mathbf{H}_\mathrm{B} \mathbf{A}_k^{\!-1}\mathbf{x} \, + \mathbf{A}_k\mathbf{w}_\mathrm{B}\in \mathbb{C}^{N \times 1}, \label{eq:y_bob} \\
\mathbf{y}_\mathrm{E} &\triangleq \mathbf{A}_{k'} \mathbf{r}_\mathrm{E} = \mathbf{A}_{k'} \mathbf{H}_\mathrm{E} \mathbf{A}_k^{\!-1} \mathbf{x} + \mathbf{A}_{k'}\mathbf{w}_\mathrm{E} \in \mathbb{C}^{N \times 1}, \label{eq:y_eve}
\end{align}
\end{subequations}
where the corresponding \ac{ML}-detected symbols are given by
\begin{subequations}
\begin{align}
\tilde{\mathbf{x}}^{\mathrm{ML}}_\mathrm{B} &\triangleq \underset{\mathbf{x} \in \mathcal{X}^{N \!\times\! 1}}{\mathrm{argmin}} \big|\big| \mathbf{y}_\mathrm{B} - \mathbf{A}_k\mathbf{H}_{\mathrm{B}}\mathbf{A}_{k}^{\!-1}\mathbf{x} \;\!\big|\big|_2^2 \in \mathbb{C}^{N \times 1}, \label{eq:x_ML_bob} \\
\tilde{\mathbf{x}}_\mathrm{E}^{\mathrm{ML}} &\triangleq \underset{\mathbf{x} \in \mathcal{X}^{N \!\times\! 1}}{\mathrm{argmin}} \big|\big| \mathbf{y}_\mathrm{E} -\! \mathbf{A}_{k'}\mathbf{H}_{\mathrm{E}}\mathbf{A}_{k'}^{\!-1}\mathbf{x} \big|\big|_2^2 \in \mathbb{C}^{N \times 1}. \label{eq:x_ML_eve}
\end{align}
\end{subequations}

Alternatively, the \ac{MMSE} equalized symbols can be obtained by
\begin{subequations}
\begin{align}
\tilde{\mathbf{x}}_\mathrm{B}^{\mathrm{MMSE}} &\triangleq \mathbf{A}_k \big(\mathbf{H}_\mathrm{B}\herm(\mathbf{H}_\mathrm{B} \mathbf{H}_\mathrm{B}\herm + \sigma^2_\mathrm{B}\mathbf{I}_N)^{-1}\big) \mathbf{r}_\mathrm{B} \in \mathbb{C}^{N \times 1}, \label{eq:x_MMSE_bob} \\[1ex]
\tilde{\mathbf{x}}_\mathrm{E}^{\mathrm{MMSE}} &\triangleq \!\mathbf{A}_{k'} \!\;\!\big(\mathbf{H}_\mathrm{E}\herm(\mathbf{H}_\mathrm{E} \mathbf{H}_\mathrm{E}\herm + \sigma^2_\mathrm{E}\mathbf{I}_N)^{-1}\big) \mathbf{r}_\mathrm{E} \in \mathbb{C}^{N \times 1}. \label{eq:x_MMSE_eve}
\end{align}
\end{subequations}

In addition, we will also consider a special ideal case for Eve, where it is considered to be co-located with Bob, such that Eve has the same \ac{CSI} as Bob, \textit{i.e.,} $\mathbf{H}_\mathrm{E} = \mathbf{H}_\mathrm{B}$ and $\sigma^2_E = \sigma^2_B$, and therefore assumed to receive the exact same signal as Bob.
%
{\color{black} In this case, the only difference between the two systems in equation \eqref{eq:r_both} are the permutation keys $k$ and $k'$, which is the only information that Eve does not have.
Nevertheless, as illustrated in Figure 1, the effective channel of Bob (matched) and Eve (mis-matched) still consist of vastly different effective channel coefficients, hence distinct demodulated signals, such that correct data detection is not possible for Eve.}


\section{Security Analysis}
\label{sec:performance}

{\color{black} In this section, we demonstrate that the proposed \acf{CP} \ac{AFDM}-based communication scheme enables ``\textit{virtually perfect}'' security, in the sense that it is virtually impossible ($i.e.$, probabilistically infeasible) for an eavesdropper to successfully decode the transmitted data without the correct permutation key, even if quantum computers are utilized.

The security of the proposed scheme will be proved by considering two key metrics at the eavesdropper Eve: 
\begin{itemize}
\item[a)] the computational hardness for the eavesdropper to determine the correct permutation key through exhaustive search, under classical and quantum computers, \vspace{0.5ex}
\item[b)] the likelihood of employing a random guess strategy to determine the correct permutation key, and the impact of close-guess cases on the data detection performance of the eavesdropper, in terms of the \ac{BER},
\end{itemize}
where both of these approaches are shown to be infeasible for the eavesdropper to correctly detect the transmitted data.
}

\subsection{Eavesdropper with Exhaustive Search Strategy}
\label{subsec:search}

\subsubsection{Classical Computing Eavesdropper} 

Given a system with $N$ number of subcarriers, there trivially exist $N!$ number of possible permutation keys in the \ac{CP}-\ac{AFDM} scheme following \eqref{eq:DAFT_matrices} and \eqref{eq:chirp_permutation}.
Consequently, an eavesdropper employing an exhaustive search must evaluate all $N!$ possible permutations to determine the correct permutation key $k \in \{1,\ldots,N!\}$.

{\color{black}
Assuming that the complexity order per metric evaluation (\ac{MMSE}, \ac{ML}, etc.) is $\mathcal{O}(\lambda)$, the total required complexity is given by $\mathcal{O}\big({\lambda \cdot (N!)}\big)$, which is clearly dominated by $N!$ for even small values of $N$.}
For example, even a very small system with $N \!=\!16$ subcarriers yields $16! \approx 2.1\!\times\! 10^{13}$, further becoming exponentially large and infeasible for larger practical systems.
{\color{black}Current \ac{5GNR} \ac{OFDM} systems utilize up to $N = 3300$ subcarriers, with larger numbers in the tens of thousands expected in \ac{6G} as systems move to higher frequencies \cite{Wang_CST23}.
Therefore, the proposed scheme is practically perfectly secure against classical computing-enabled eavesdroppers under exhaustive search attacks.
}

\vspace{1ex}

\subsubsection{Quantum Computing Eavesdropper} Let us now consider an eavesdropper equipped with a quantum computer,
{\color{black}
which are known to perform efficient \ac{ML} searches with quadratic acceleration by means of the \ac{GAS} algorithm \cite{Gilliam_Quantum21,Yukiyoshi_VTC24,Sano_TQE24}, at the cost of at least $\log_2(N!)\approx N\log_{2}(N)-N\log_{2}(e)+\mathcal{O}\big(\log_{2}(N)\big)$ logical qubits, where the latter expression is the well-known Stirling's formula.

Since quantum computers typically require, for each logical qubit, $10^{3}\!\sim\!10^{4}$ as much additional qubits for error correction, the latter implies that a quantum computer with approximately $\log_2(3300!) \approx 3.5\times 10^{7}\!\sim\!10^{8}$ qubits would be needed in order to perform a \ac{GAS}-based search over all possible permutations in a proposed \ac{CP}-\ac{AFDM} system with $N=3300$ subcarriers, which is $10$ to $100$ times more than the $10^6$ qubits expected to equip quantum computers by 2040 \cite{Choi_Arxiv23}.
To make matters worse, the required query complexity associated with the \ac{GAS} algorithm is of order $\mathcal{O}\big(\sqrt{N!}\,\big)$ \cite{Gilliam_Quantum21}, such that for $N=3300$, the search performed by Eve's quantum computer would have an overwhelming query complexity of order $\mathcal{O}\big(\sqrt{N!}\big)\approx \mathcal{O}\big((2\pi N)^{\!\frac{1}{4}}\!\left({\frac{N}{e}}\right){\vphantom{x}}^{\!\!\frac{N}{2}}\big)\approx \mathcal{O}(1.8\times 10^{1654})$.

It follows that the proposed \ac{CP}-\ac{AFDM} scheme, if employed over systems with $N=3300$ subcarriers, is quantum resilient at least up to 2040, with the resilience of systems employing larger numbers of subcarriers going farther beyond.}

\vspace{-2ex}
\subsection{Eavesdropper with Random Guess Strategy}
\label{subsec:randomGuess}
\vspace{-0.25ex}

As the exhaustive search strategy has been shown to be infeasible even via quantum computers, next, we consider the case where the eavesdropper employs a random guess strategy {\color{black}to utilize a random permutation sequence for data detection.}

\vspace{1ex}

\subsubsection{Permutation Guessing Probability}
\label{sec:probability}

{\color{black}In this case, the metric used to measure the security of the method is the likelihood that Eve blindly guesses} the correct permutation order $k${\color{black}, which is straightforwardly given by} $\frac{1}{N!}$.
{\color{black}Again, such probability is found to be virtually zero even for relatively small $N$, due to the factorial term in the denominator.}

%

\begin{figure}[H]
\centering
\begin{subfigure}[h]{1\columnwidth}
\includegraphics[width=1\columnwidth]{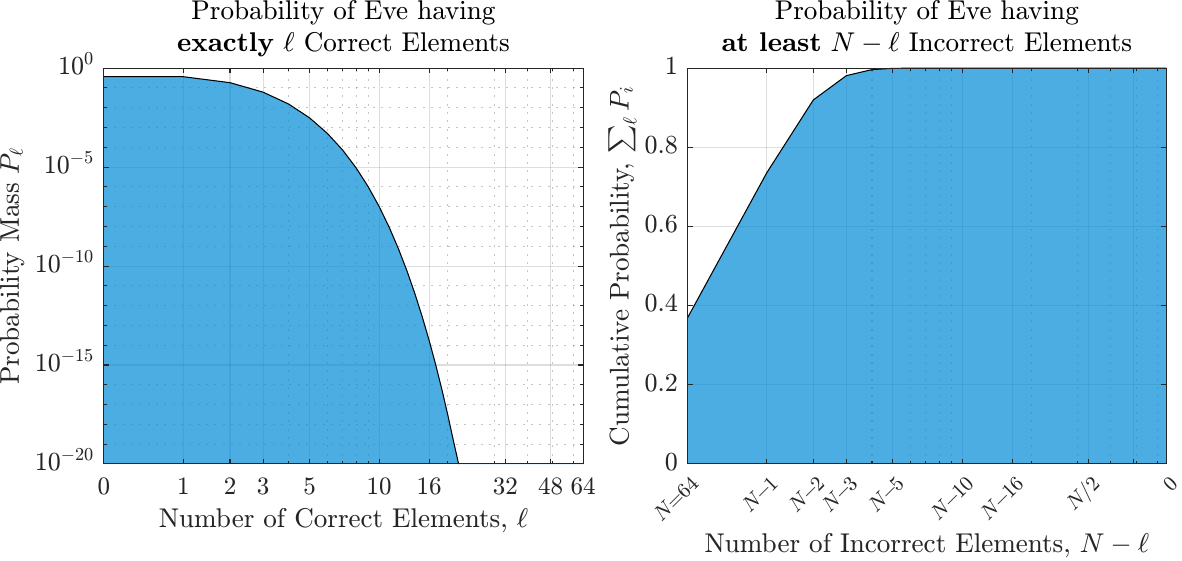}
\vspace{-3.5ex}
\caption{$N = 64$ (i.e., small Wi-Fi system).}
\label{fig:PMFCDF_N64}
\end{subfigure}%
\vspace{1.5ex}

\begin{subfigure}[h]{1\columnwidth}
\includegraphics[width=1\columnwidth]{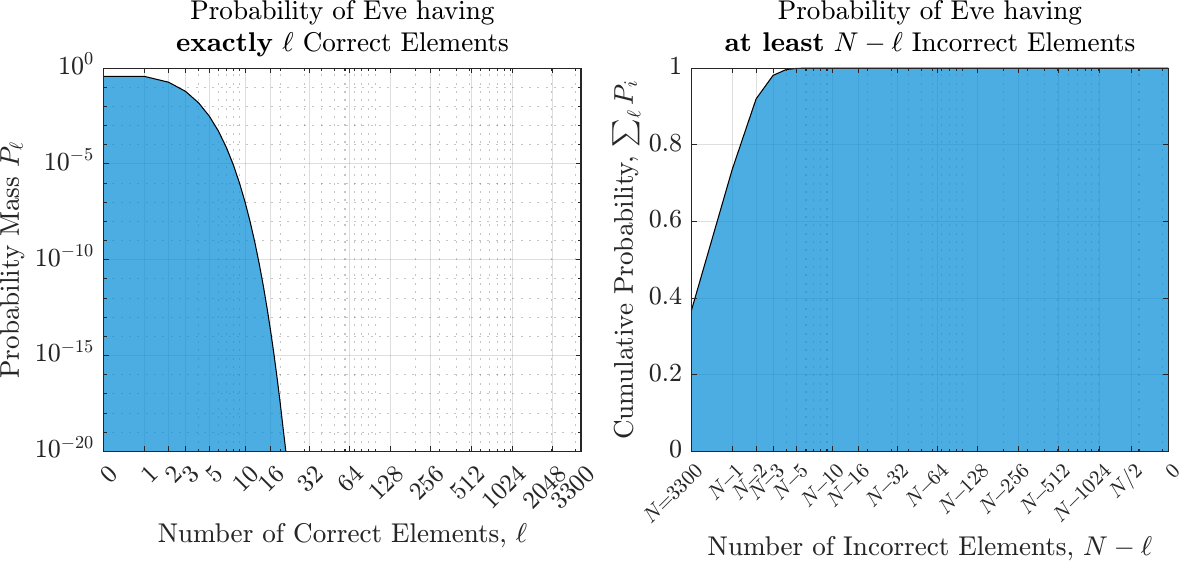}
\vspace{-3.5ex}
\caption{$N = 3300$ (i.e., large \ac{5GNR} system).}
\label{fig:PMFCDF_N3300}
\end{subfigure}
\caption{\color{black}Probabilities of Eve guessing a sequence which is $\ell$-elements close to the true sequence, in terms of the \acs{PMF} of the number of correct elements $\ell$, and the \acs{CDF} of the number of incorrect elements $N-\ell$.}
\label{fig:PMFCDF}
\vspace{-1ex}
\end{figure}

{\color{black}
The latter result is only a particular case of the more general problem of determining the likelihood that Eve makes} a random guess close to the true permutation.
{\color{black}To elaborate, let us suppose that Eve selects a permutation order $k' \in \{1, \ldots, N!\}$, such that there are $\ell \in \{0,1,\ldots,N\}$ number of elements in Eve's permuted sequence matching the positions of the corresponding elements in the \textit{correct} permuted sequence with order $k$ (of Alice).
In other words, if $\ell = 0$, Eve's sequence has \ul{no} elements that are in their correct positions with respect to the true sequence, and conversely, $\ell = N$ indicates that all $N$ elements are correctly positioned.}

Leveraging combinatorial analysis, {\color{black}the total number of permutations with $\ell$ correct element positions}, out of $N!$ possible sequences, can be obtained by considering derangements \cite{Hassani_2003}, from which the {\color{black}corresponding} probability can be obtained as \vspace{-0.5ex}
\begin{equation}
    \color{black}
P_{\ell} \triangleq \frac{D_{N-\ell}}{N!} = \frac{1}{\,\ell!\,}\sum_{n = 0}^{N - \ell} \frac{(-1)^n}{n!}, \vspace{-0.5ex}
\end{equation}
{\color{black}
where $D_{n}$ is the number of size-$n$ derangements \cite{Hassani_2003}, and $P_{\ell}$ is the probability that Eve's randomly guessed sequence contains $\ell$ correctly positioned elements\footnote{\color{black}Since any permutation of a sequence involves at least two elements, permuted sequences with a single unmatched element does not exist, i.e., $P_{\ell =  N \!-\! 1} = 0$.}, which confirms the previous trivial case with all correct elements, $P_{\ell =N} = \tfrac{1}{N!}$.

Figure \ref{fig:PMFCDF} illustrates the {\color{black} \acp{PMF} and \acp{CDF} for practical number of subcarriers, of $N = 64$ (i.e., small Wi-Fi system) and $N = 3300$ (i.e., large \ac{5GNR} system)}.
The \acp{PMF} reveal the exponentially decreasing probability of an eavesdropper correctly guessing a permutation with a {\color{black} number $\ell$ of elements in the correct positions}, where the probabilities quickly drop to negligible values even for small values of $\ell$.

The corresponding \acp{CDF} are also shown, with the $x$-axes redefined in terms of the probability that Eve guesses a permutation with at least $(N - \ell)$ incorrect elements, providing a more intuitive interpretation of the results, by highlighting that the eavesdropper is practically bound to guess permutations with mostly wrong elements, where the \ac{CDF} reaches unity (certainty) with only $\ell = 5 \sim 10$, for both system sizes.

In fact, it is known \cite{Hassani_2003} that for large $N$, the derangement probabilities $P_{\ell}$ converge to values that are independent of $N$, $i.e$., $P_{\ell = 0} \rightarrow \tfrac{1}{e}$, such that \ac{CDF} for the first several values of $\ell$ are identical for systems of any (sufficiently large) size.
For example, it is found that Eve is bound to guess a permutation with at least $N - 10$ incorrect elements, in other words, a permutation with \textit{no more} than $\ell = 10$ correct entries, with a near-certain probability of $99.999999992\%$, which implies that in a system with $N >\!\!> 10$ subcarriers, the sequence obtained by Eve will be virtually sure to be far different from the true.}

\subsubsection{BER Performance Analysis}

\begin{figure}[b]
    \vspace{-3ex}
\centering
\includegraphics[width=1\columnwidth]{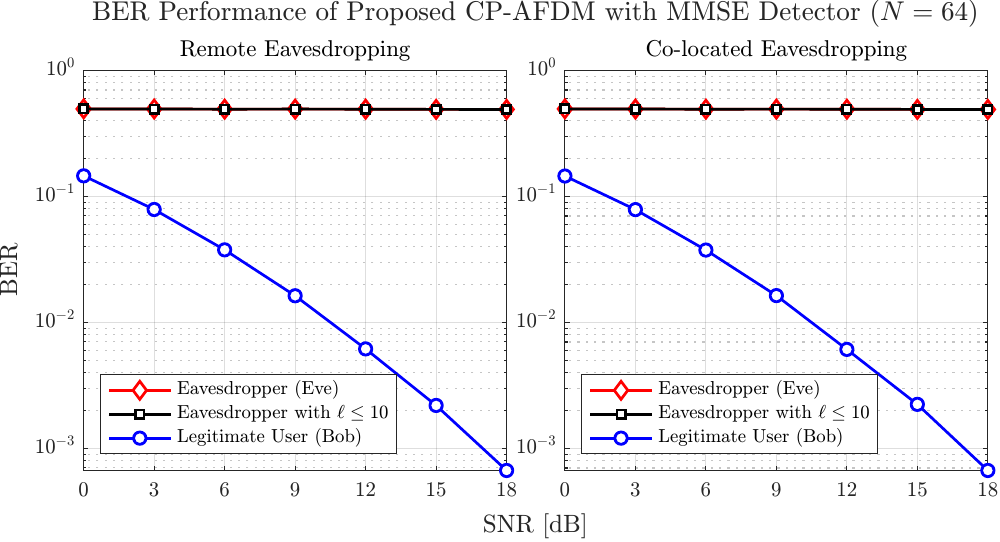}
\caption{BER performance of Bob and Eve, {\color{black}and a special case of $\ell \leq 10$, in a remote (left) and co-located (right) eavesdropping scenario with $N = 64$}.}
\label{fig:BER}
\end{figure}

{\color{black} Following the above analysis which shows that the eavesdropper will, with practical certaintly, obtain a permutation sequence vastly different to that of the legitimate users, we close the full circle by showing that the corresponding signal demodulation and data detection is not possible in both remote and co-located eavesdropping scenarios, enabling the proposed secure communications scheme.

Figure \ref{fig:BER} illustrates the \ac{BER} performance obtained via numerical simulations, of the eavesdropper Eve and the legitimate user Bob, where the \ac{MMSE} estimator for data detection is constructed using the random permutation key $k'$ by Eve, and using the correct permutation key $k$ by Bob.}
Two sets of numerical results are presented, for the remote eavesdropping {\color{black}($\mathbf{H}_\mathrm{E} \neq \mathbf{H}_\mathrm{B}$)} and co-located eavesdropping {\color{black}($\mathbf{H}_\mathrm{E} = \mathbf{H}_\mathrm{B}$)} scenarios respectively, {\color{black}in a system with $N = 64$ subcarriers.}




{\color{black}
The illustrated results clearly show that in both the remote and co-located eavesdropping scenarios, the proposed scheme enables virtually perfect security against Eve with the random guessing approach, who cannot detect any information as shown by the plot in red.}
{\color{black}In addition, the identical \ac{BER} performances for both Bob and Eve between the two scenarios highlight that the security of the proposed scheme is only dependent on the novel chirp-permutation effect of the proposed \ac{CP}-\ac{AFDM} waveform, and not on the physical channel.}

{\color{black} Finally, a special analytical case of the eavesdropper is shown, which only considers the \ac{BER} performance under permutations with $\ell \leq 10$, shown previously to occur with near-certainty.
The analysis of this case, which therefore dominates the true random guessing scenario, clearly indicates that no successful data detection is possible with $\ell \leq 10$, i.e., the received \ac{CP}-\ac{AFDM} signal is not decodable with permutations with no more than $\ell = 10$ correct elements, in any scenario.}

\vspace{-1ex}
\section{Conclusion}
\label{sec:conclusion}
\vspace{-0.5ex}

We proposed a novel secure communications scheme based on the chirp-permuted \ac{AFDM} waveform, which provides quantum-resilient physical-layer security against eavesdroppers in a doubly-dispersive channels.
{\color{black} It is shown that it is practically infeasible to detect correct information, for eavesdroppers without the secret permutation-order key used by the transmitter, under random guess strategies, and the classical or quantum computing-based exhaustive search strategies.}

\vspace{-1ex}
\section{Acknowledgement}
\label{sec:ack}

A part of this work was conducted as part of Project ``5G-HyprMesh" (Code 01MO24001C), funded by the Bundesamt für Sicherheit in der Informationstechnik (BSI).



\vspace{-1ex}

\vspace{-0.25ex}

\end{document}